\documentclass[journal]{IEEEtran}

\usepackage[utf8]{inputenc}
\usepackage{graphicx} 
\usepackage{cite}
\usepackage{amsmath,amssymb,amsfonts}
\usepackage{algorithm}
\usepackage{algpseudocode}
\usepackage{array}
\usepackage{booktabs}
\usepackage{makecell}
\usepackage{url}
\usepackage{balance}
\usepackage{xcolor}
\usepackage{pifont}
\usepackage{enumitem}
\usepackage{tabularx}
\usepackage{mathrsfs}
\usepackage{placeins}
\usepackage{float}
\usepackage{bm}

\title{\LARGE Phantom of the Masquerade: Graph Neural Network-Based Detection of Masquerade Attacks in Controller Area Networks}
\author{
Nehal Ameen$^{*}$, \textit{Member, IEEE},
Pablo Moriano$^{\dagger}$, \textit{Senior Member, IEEE},
Irfan Ahmed$^{*}$, \textit{Senior Member, IEEE}
\\
$^{*}$Department of Computer Science, Virginia Commonwealth University, Richmond, VA 23284, USA
\\
$^{\dagger}$Computer Science and Mathematics Division, Oak Ridge National Laboratory, Oak Ridge, TN 37830, USA
\\
$^{*}$\{ameennm, iahmed3\}@vcu.edu, $^{\dagger}$moriano@ornl.gov
}

\date{}

\begin{document}

\maketitle

\begin{abstract}

The lack of built-in security in controller area networks (CANs)
exposes modern vehicles and industrial systems to masquerade
attacks that can compromise safety and reliability. These attacks
inject malicious messages that mimic legitimate communication
patterns, making them difficult to detect using existing intrusion
detection systems (IDS). Prior graph-based IDS approaches often
rely on shallow embeddings that decouple representation learning
from classification, which may limit their ability to fully exploit
structural and signal-level information. In this paper, we propose
PhantoMasq, a graph neural network (GNN)-based framework for
detecting masquerade attacks in CAN systems. PhantoMasq models
CAN traffic as message sequence graphs enriched with lightweight
statistical features derived from signal behavior, enabling joint
learning of structural dependencies and temporal characteristics.
The detection task is formulated as a graph-level classification
problem and solved using graph convolutional networks (GCNs).
We evaluate PhantoMasq on the ROAD dataset
under a protocol that prevents overlapping windows from spanning
training and test data. Across five masquerade attack types,
PhantoMasq achieves mean AUC-ROC between 0.831 and 0.999, with
performance depending on the transmission rate of the targeted
identifier and the magnitude of the induced signal deviation.
Against a state-of-the-art graph-embedding baseline evaluated
on the same sliding-window configurations, PhantoMasq achieves
higher AUC-ROC on all five attack types, by 0.13 to 0.36, and
is less sensitive to window configuration. Window-processing
latency ranges from 2.33 to 5.94 ms per window on CPU,
corresponding to a throughput of 168 to 429 windows per second.
We make our implementation and experimental artifacts available
to support reproducibility.

\end{abstract}

\begin{IEEEkeywords}
Controller area network (CAN), intrusion detection system (IDS), graph neural network (GNN), masquerade attacks.
\end{IEEEkeywords}

\section{Introduction}
Modern vehicles are evolving into complex, connected cyber-physical systems, deeply reliant on the controller area network (CAN) protocol for real-time communication between electronic control units (ECUs). CAN underpins critical systems across not only automotive platforms, but also industrial automation and medical devices \cite{hpl2002introduction}. Its cost-effectiveness and low wiring overhead have made CAN a foundational technology. However, despite its widespread adoption, CAN lacks built-in security mechanisms, making it critically vulnerable to cyberattacks. Similar threats affect embedded controllers in other cyber-physical systems, such as programmable logic controllers in industrial control systems, where attacks and detection methods have both been studied~\cite{ameen2025wavesleuth, kalle2019clik}.

Over the past decade, a series of high-profile attacks have demonstrated the real-world consequences of insecure CAN deployments. In 2015, two researchers remotely gained physical control of a Jeep Cherokee, prompting Fiat Chrysler to recall 1.4 million vehicles \cite{miller2015remote, bbc2015jeep}. Shortly after, researchers demonstrated remote brake disablement in a Corvette via CAN message injection \cite{191966}. Similar vulnerabilities were exposed in a Tesla Model S \cite{guardian2016tesla} and in heavy commercial vehicles \cite{burakova2016truck}. More recently, CAN injection attacks have been used in coordinated vehicle thefts, affecting not only passenger vehicles but also critical fleets such as ambulances and buses \cite{securityweek2023can}. These incidents underscore the urgent need for robust CAN security mechanisms.

Originally developed in 1983 by Robert Bosch GmbH~\cite{farsi1999overview, johansson2005vehicle}, CAN was never designed with adversarial environments in mind. Its lack of encryption compromises confidentiality, while integrity is not guaranteed because error-detection mechanisms cannot distinguish malicious manipulation from benign faults. Availability is also at risk due to arbitration rules that allow high-priority nodes to dominate communication, enabling denial-of-service attacks \cite{moriano2022detecting}.

Among the most challenging threats are masquerade attacks, in which adversaries inject syntactically valid CAN messages that mimic legitimate traffic patterns \cite{jo2019mauth}. These attacks can trigger unintended acceleration, disable braking systems, or manipulate steering, while evading traditional intrusion detection systems (IDS) that rely on rule-based or shallow learning techniques \cite{sharmin2023comparative}. Their ability to preserve expected communication frequencies makes them particularly difficult to detect.

Several deep learning-based IDS frameworks have emerged, including sequential models such as long short-term memory (LSTM) and gated recurring unit (GRU) architectures \cite{cahuantzi2023comparison, pudikov2020comparison}, as well as graph-based approaches \cite{yang2024anomaly, du2025ugl}. However, many approaches do not jointly model temporal dynamics and graph structure within a unified and reproducible graph-level framework, often treating messages independently or relying on heavily engineered features~\cite{Shahriar:2023:CANShield, shahriar2023cantropy}. Moreover, many studies rely on proprietary datasets or closed-source implementations, hindering reproducibility and limiting real-world applicability~\cite{he2024d, el2024safeguarding}.

In this paper, we propose PhantoMasq, a graph neural network (GNN)–based framework for detecting masquerade attacks in CAN systems. Our approach represents CAN traffic as message sequence graphs (MSGs) that capture both communication structure and temporal behavior through lightweight statistical features derived from CAN signals. This graph-based representation enables learning from
sequential relationships among CAN message identifiers,
allowing the model to identify subtle deviations associated
with masquerade attacks. We formulate detection as a graph-level classification problem and use graph convolutional networks (GCNs) to distinguish between benign and malicious traffic.

We evaluate PhantoMasq on the publicly available Real ORNL Automotive Dynamometer (ROAD) dataset \cite{verma2024comprehensive}, a widely used benchmark in vehicular network security, under a protocol that splits each capture temporally before windowing, so that overlapping windows never span training and test data. Across five masquerade attack types, PhantoMasq achieves mean AUC-ROC between 0.831 and 0.999. Attacks that induce large deviations on frequently transmitted identifiers are detected almost perfectly, while performance is lower for attacks that alter a single bit or target a sparsely transmitted identifier. Compared with a state-of-the-art graph-embedding baseline evaluated on the same sliding-window configurations, PhantoMasq achieves higher AUC-ROC on all five attack types, with a window-processing latency of 2.33--5.94 ms
per window on CPU. In addition, we make the PhantoMasq implementation
and experimental artifacts available to support
reproducibility~\cite{nehal2023gnns_can}. The main contributions of this paper are summarized as follows:

\begin{itemize}
\item We introduce a graph-based representation of CAN traffic using message sequence graphs that capture both structural interactions and temporal behavior.
\item We formulate masquerade attack detection as a graph-level classification problem and apply GNNs to learn representations of CAN communication patterns.
\item We demonstrate that lightweight statistical features support efficient detection of masquerade attacks, and characterize how detection difficulty depends on the transmission rate of the targeted identifier and the magnitude of the induced
signal deviation.
\item We evaluate under a protocol that prevents leakage between overlapping windows, report AUC-PR alongside AUC-ROC to account for class prevalence as low as 5.4\%, and provide the implementation and experimental artifacts to support reproducibility.
\end{itemize}

\section{Background}
\subsection{CAN Protocol}

CAN is a widely adopted and robust communication protocol that facilitates the exchange of messages between ECUs in automotive and industrial applications, as well as medical devices \cite{di2012understanding}. The CAN bus protocol was initially developed by Robert Bosch GmbH, and the CAN 2.0 specification was released
in 1991 \cite{farsi1999overview,johansson2005vehicle},
revolutionizing automotive communication technologies. CAN was further standardized internationally in 2011 by ISO 11898-1:2011 \cite{boland2021overview}. That standardization specified the requirements for high-speed, light-weight communication and data exchange among various vehicle subsystems, without taking security into account. 

CAN is a broadcast protocol that operates on a multi-master serial bus standard, enabling all ECUs connected to the bus to send and receive signals at high speeds, up to 1 Mbps, without a central computer. CAN frames carry messages containing sensor readings,
control information, and other data exchanged among ECUs. For example, ECUs such as the transmission control module, engine control unit, and the anti-lock braking system broadcast data frames that provide information about the vehicle's current state \cite{halder2020coids,shahriar2023cantropy}. 

CAN encompasses the first two layers of the Open Systems Interconnection (OSI) model, namely the physical and data-link layers \cite{hpl2002introduction}. A CAN frame comprises several fields, two of which we focus on for our research. The first is the identifier (ID) field that serves as the frame header. This unique identifier represents the message priority, as it is responsible for arbitration, the process of prioritizing frames when multiple ECUs are transmitting signals simultaneously. Lower values indicate higher priorities. Standard IDs consist of 11 bits, while extended IDs consist of 29 bits. The second field we focus on is the payload, or data, field. It contains the actual data carried by the frame, such as sensor readings or control commands, encapsulated in a data field ranging from 0 to 8 bytes. Each unique piece of information carried in the message is a distinct signal. CAN frames that have the same ID encode the same set of signals with the same format and are typically transmitted at a fixed frequency to relay updated signal values.

To extract these signals from CAN payloads, a CAN
database (DBC) file is used to decode the binary data
into interpretable signal values, providing time-series
information on the network's activity
\cite{shen2025dbcsec}. This process produces a structured,
time-series representation of CAN signals, which is commonly
used in data-driven intrusion detection systems to analyze
relationships between physical signals.

The attacks that CAN intrusion detection systems aim to
detect can be broadly classified into three categories
\cite{verma2024comprehensive}:

\begin{itemize}
    \item \textbf{Fabrication attacks:} A rogue ECU injects malicious messages into the network, without disrupting normal traffic from other ECUs \cite{gazdag2023crysys}.
    \item \textbf{Suspension attacks:} A legitimate ECU gets disabled or disconnected, causing it to stop sending or receiving messages temporarily \cite{al2024can}.
    \item \textbf{Masquerade attacks:} A legitimate ECU is first suppressed by a weak attacker, after which a strong attacker impersonates it by injecting spoofed frames at realistic timing and signal patterns \cite{moriano2022detecting}.
\end{itemize}
Masquerade attacks are particularly challenging to detect due to their ability to mimic legitimate traffic patterns. These structural and temporal characteristics of CAN traffic motivate graph-based representations, which can capture interactions between message identifiers and support learning-based intrusion detection.

\subsection{GNNs}

GNNs are neural networks designed to operate on graph-structured data, where entities are represented as nodes and relationships as edges. GNNs have been widely applied in domains such as cybersecurity and network analysis \cite{zhong2024survey}.

GNNs learn representations of nodes or entire graphs by iteratively aggregating information from neighboring nodes through a message-passing mechanism, enabling the model to capture structural dependencies and interactions within the data. In particular, GNNs can be applied to graph-level classification tasks, where the goal is to assign a label to an entire graph based on its structure and node features \cite{hamilton2020graph}.

Among GNN architectures, GCNs are a widely used class of models that leverage node features and local graph connectivity to learn effective representations for classification tasks, while maintaining computational efficiency and strong empirical performance \cite{wu2019simplifying}, making them suitable for graph-level intrusion detection in CAN networks.

\section{Threat Model and Problem Definition}

Masquerade attacks in CAN pose significant risks to system safety and reliability. In this section, we formalize the system assumptions, adversarial capabilities, and detection objectives considered in this work.

\subsection{System Model}

We consider an in-vehicle network based on the CAN protocol, where multiple ECUs communicate via broadcast messages over a shared bus. Each CAN message is identified by an ID and carries payload data corresponding to signal transmissions between the ECUs.

We assume that CAN traffic is continuously monitored and collected at a central observation point (e.g., gateway ECU or logging module), without modifying the underlying CAN protocol. The monitoring system observes sequences of CAN messages and processes them over fixed-length segments of consecutive CAN frames, which are later implemented as sliding windows in our methodology.

Within each segment, CAN messages are transformed into a graph representation that captures message interactions over time. Node attributes are derived from lightweight statistical features computed from decoded CAN payloads. The detection system operates passively and does not interfere with the normal operation of the CAN bus.

\subsection{Adversary Model}

We consider a capable adversary with access to the CAN bus, either physically or remotely through a compromised ECU. The adversary can inject malicious CAN messages to manipulate vehicle behavior.

We focus specifically on masquerade attacks, which are among the most sophisticated and stealthy attack types in CAN. These attacks combine elements of fabrication and suspension. A \textit{weak attacker} learns the transmission frequency of messages sent from a targeted ECU and suppresses its legitimate transmissions, while a \textit{strong attacker} injects malicious messages at the expected frequency using the same ID. By preserving expected timing patterns and identifiers,
these attacks can evade detection methods that rely
primarily on these characteristics.

The adversary is assumed to have knowledge of:
\begin{itemize}
    \item valid message identifiers,
    \item typical message transmission frequencies,
    \item general communication patterns on the CAN bus.
\end{itemize}
However, we assume that the adversary does not have complete control over all ECUs simultaneously and cannot perfectly replicate the joint statistical and structural dependencies present in legitimate CAN traffic over time. We do not assume access to cryptographic protections either, as CAN lacks built-in authentication mechanisms.

This threat model aligns with commonly studied CAN intrusion scenarios and reflects realistic attacker capabilities in vehicular networks \cite{marfo2025detecting}.

\subsection{Detection Objective}

Given a sequence of CAN messages segmented into fixed-length segments of consecutive CAN frames, the objective is to classify each segment as benign or malicious (masquerade attack) behavior.

For each segment, the detection system constructs a graph representation and performs graph-level classification to label it as benign or malicious.

The detection model should satisfy the following goals:
\begin{itemize}
    \item \textbf{Accuracy:} achieve high detection performance across diverse masquerade attack scenarios.
    \item \textbf{Robustness:} maintain detection performance across masquerade attack types and sliding-window configurations.
    \item \textbf{Efficiency:} operate with low latency to support near real-time deployment.
\end{itemize}

\section{Related Work}

We discuss prior work closely related to GNN-based CAN IDS in Section~\ref{subsec:Closely Related Work}, followed by other graph-based CAN IDS approaches in Section~\ref{subsec:Other Related Work}.

\subsection{Closely Related Work} \label{subsec:Closely Related Work}

Recent research on IDS for CAN bus networks has increasingly turned to GNN-based approaches to capture the complex dependencies among messages exchanged between ECUs.

Zhang et al. \cite{zhang2023federated} generated directed attributed graphs based on CAN message streams over fixed intervals. Message contents were encoded as node attributes, while the frequency of ID pairs was captured as edge attributes. They proposed a two-stage classifier combining one-class anomaly detection with multi-class attack classification, and incorporated federated learning to preserve data privacy. Their system detects multiple attack types within 3 milliseconds.

Yang et al. \cite{yang2024anomaly} introduced CAN-ASTGCN, which detects anomalies by modeling correlations among CAN bus sensors. Their approach combines an attention-based spatio-temporal GCN with a Sensor Parameter Adaptive Learning module, achieving 97.68\% accuracy across multiple attack scenarios.

He et al. \cite{he2024d} proposed the A\&D-GNN, which constructs directed attributed graphs using both arbitration and payload fields to capture richer context, achieving 99.92\% accuracy on impersonation attacks in the OTIDS dataset. El-Dawy et al. \cite{el2024safeguarding} developed a GAT-based IDS where nodes represent IDs and edges encode message sequences, achieving real-time detection with latency around 7 ms and up to 100\% accuracy, while analyzing the trade-off between graph size and runtime.

Wang et al. \cite{wang2025dynamic} introduced a dynamic semi-supervised IDS that constructs temporal graphs and applies a Temporal Graph Attention Network (TGAT) with a TimeMemory module. Their method leverages unlabeled data and supervised contrastive learning, achieving very high F1-scores on the Car-Hacking dataset.

Gao et al. \cite{gao2025signal} modeled CAN data as dynamic multivariate signal graphs using graph transformers. Signals were treated as nodes and dependencies as edges, with node features obtained through reverse engineering of CAN payloads. Their method was evaluated on the Car-Hacking and CAN-Injection datasets.

Du et al. \cite{du2025ugl} proposed UGL, a hybrid GCN and LSTM model designed for UAVCAN systems. Their approach emphasizes graph construction and feature engineering, incorporating message cycle, Hamming distance, and entropy, achieving up to 100\% detection accuracy for flooding attacks.

Hasan et al. \cite{hasan2025can} introduced CAN-GraphiT, a transformer-based IDS that jointly models temporal and graph features using self-attention. Their method achieves up to 98.66\% accuracy for masquerade attacks and maintains low inference latency (0.1 ms), outperforming several deep learning baselines.

Jiang et al. \cite{jiang2025xiphos} proposed an unsupervised graph contrastive learning framework combined with adaptive hierarchical clustering. While promising, their method achieved an AUC of 75\% on the ROAD dataset.

Table~\ref{tab:state-of-the-art} summarizes these GNN-based
CAN intrusion detection approaches in terms of problem
setting, model architecture, datasets, evaluation metrics,
and code/data availability.

\textit{Contrast with our work:}  
Prior GNN-based approaches often rely on complex temporal encoders, attention mechanisms, or specialized graph constructions. In contrast, our work adopts a simpler formulation by casting masquerade detection as a graph-level classification problem and using a GCN to learn representations directly from annotated MSGs. By combining structural information with lightweight statistical features, our approach achieves mean AUC-ROC scores ranging from 0.831 to 0.999 across five ROAD masquerade attack types, using a relatively simple GCN architecture. Our work specifically targets masquerade attacks, evaluates computational efficiency, and releases the implementation and experimental artifacts to support reproducibility.

\newcommand{\cmark}{\textcolor{green}{\ding{51}}}
\newcommand{\xmark}{\textcolor{black}{\ding{55}}}
\newcolumntype{C}{>{\centering\arraybackslash}X}

\begin{table*}[ht]
    \centering
    \caption{Comparison of State-of-the-Art on GNN-based IDS in CANs.}
    \label{tab:state-of-the-art}
    \setlength{\tabcolsep}{3pt}
    \renewcommand{\arraystretch}{1.2}
    
    \begin{tabularx}{\textwidth}{l C C C C C}
         \textbf{Reference} & \textbf{Problem Setting} & \textbf{Model} & \textbf{Dataset(s)} & \textbf{Metrics} & \textbf{Code/Data} \\ \hline
         Zhang et al. (2023) \cite{zhang2023federated} & Graph-level anomaly + classification & Federated GNN & Ford Transit, HCRL & Accuracy, Recall, F1, latency & \xmark \\ \hline
         Yang et al. (2024) \cite{yang2024anomaly} & Node + graph anomaly detection & ASTGCN & Survival Dataset, Car Hacking & Accuracy, Precision, Recall, F1 & \xmark \\ \hline
         He et al. (2024) \cite{he2024d} & Graph-level detection & A\&D-GNN & OTIDS & Precision, Recall, F1 & \xmark \\ \hline
         El-Dawy et al. (2024) \cite{el2024safeguarding} & Graph-level detection & GCN + GAT & OTIDS & Accuracy, latency & \xmark \\ \hline
         Wang et al. (2025) \cite{wang2025dynamic} & Node-level detection & TGAT & Car Hacking & Accuracy, F1, AUC & \xmark \\ \hline
         Gao et al. (2025) \cite{gao2025signal} & Node + graph detection & Graph Transformer & Car Hacking, Car-Injection (self-collected) & Precision, Recall, F1 & \xmark \\ \hline
         Du et al. (2025) \cite{du2025ugl} & Node-level detection & GCN + LSTM & UAVCAN & Accuracy, Recall & \xmark \\ \hline
         Hasan et al. (2025) \cite{hasan2025can} & Graph-level detection & Transformer & Car Hacking, ROAD, CAN-MIRGU, CrySys & Accuracy, ROC, latency & \xmark \\ \hline
         Jiang et al. (2025) \cite{jiang2025xiphos} & Graph contrastive learning & GCL + AHC & Car Hacking, ROAD & AUC, F1 & \cmark \\ \hline
         \textbf{PhantoMasq (This Work)} & \textbf{Graph-level masquerade detection} & \textbf{GCN} & \textbf{ROAD} & \textbf{AUC-ROC, AUC-PR, TTW, throughput} & \cmark \\ \hline
    \end{tabularx}
\end{table*}

\subsection{Other Related Work} \label{subsec:Other Related Work}

While GNN-based IDSs have become prominent for modeling CAN traffic, several studies have explored graph-based intrusion detection frameworks that leverage structural representations without relying on graph neural architectures. These works highlight the broader utility of graph abstractions in capturing message interactions, even when learning is not performed through GNNs.

Islam et al. \cite{islam2020graph} proposed one of the earliest graph-based IDS frameworks using a chi-squared method to detect attacks, achieving low misclassification rates across multiple attack types on the HCRL dataset.

Jedh et al. \cite{jedh2021detection} used MSGs and graph-based analytics to detect malicious injections with 97.32\% accuracy and 2.5 ms latency, demonstrating the feasibility of graph-based detection without relying on proprietary information.

Refat et al. \cite{refat2022lightweight} applied neighborhood-based graph similarity for lightweight IDS design, achieving 96.01\% accuracy on real vehicle data.

Meng et al. \cite{meng2023gb} introduced GB-IDS, combining graph structures with a variational autoencoder to enable training without negative samples, achieving high detection rates on OTIDS.

Ye et al. \cite{ye2025gdt} proposed GDT-IDS using graph-derived features such as time difference, centrality, and density, achieving strong accuracy while maintaining interpretability.

Marfo et al. \cite{marfo2025detecting} combined shallow graph embeddings (node2vec) with time-series features, using ensemble classifiers such as Random Forest and XGBoost. Their method demonstrated competitive performance on masquerade attacks in the ROAD dataset.

\textit{Contrast with our work:}  
Compared to these approaches, which rely on statistical testing, similarity measures, or shallow graph embeddings, our method employs end-to-end graph representation learning through a GCN. This enables direct learning of joint structural and feature representations without requiring handcrafted graph metrics or separate classifiers. In contrast to node2vec-based pipelines, which decouple representation learning from classification, our approach integrates both stages into a unified framework. We evaluate its detection performance and computational efficiency relative to existing approaches in Section~\ref{subsec:Comparison with SOTA Method}.

\section{Methodology}
\begin{figure*} [!ht]
    \centering
    \includegraphics[width=1\linewidth]{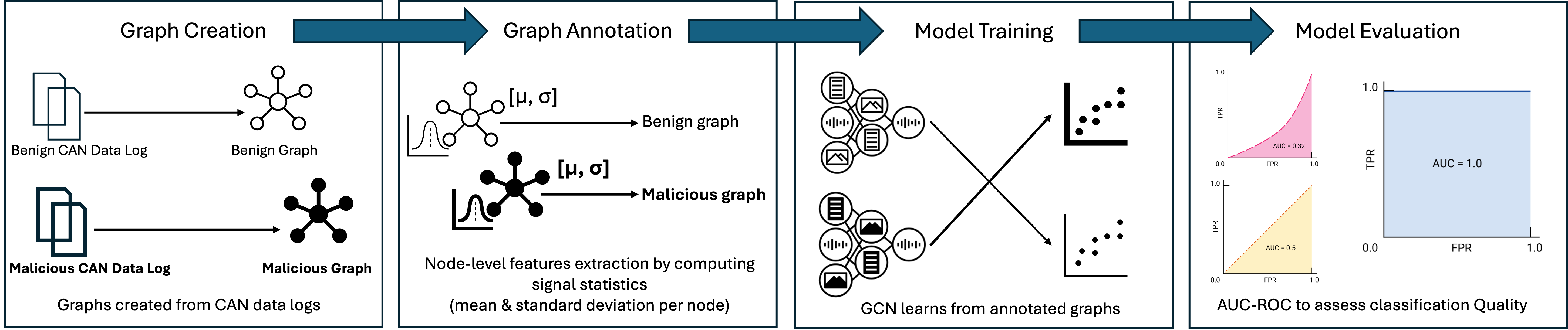}
    \caption{Overview of the PhantoMasq four-stage pipeline for graph-level CAN intrusion detection.}
    \label{fig:can-gnn-pipeline}
\end{figure*}

This section describes the methodology used to develop and evaluate the proposed GNN-based framework for detecting masquerade attacks in CAN systems. The approach defines a processing pipeline from data partitioning and graph construction to feature integration and graph-level classification. As shown in Fig. \ref{fig:can-gnn-pipeline}, the methodology consists of four stages designed for effectiveness and reproducibility.

\subsection{Methods Overview}

Our approach models masquerade attack detection as a graph-level classification problem by integrating structural information from MSGs with statistical features derived from decoded CAN signals. Rather than relying on shallow graph embeddings followed by separate classifiers, we use a GCN as an end-to-end learning framework over annotated graphs. As shown in Fig. \ref{fig:can-gnn-pipeline}, the pipeline partitions CAN data into fixed-length segments, constructs MSGs, enriches them with statistical node attributes, and applies a GCN for classification.

\subsection{CAN Data Partitioning and MSG Building}

To enable continuous analysis of CAN traffic, we partition the incoming stream of CAN messages into fixed-length segments of consecutive CAN frames, implemented as sliding windows parameterized by window size $\omega$ and offset $\delta$. Each message is represented by its timestamp, ID, and payload.

We implement this segmentation using a sliding window mechanism \cite{moriano2026evaluating}, parameterized as follows:

\begin{itemize}
    \item \textbf{Window size ($\omega$):} The length of each window, determining the number of CAN frames included in each graph representation.
    \item \textbf{Offset ($\delta$):} the number of frames between the starts of consecutive windows. Consecutive windows overlap by ($\omega - \delta$) frames when ($\delta < \omega$).
\end{itemize}

To construct labeled evaluation data, each attack capture is first partitioned into three contiguous regions using the ground-truth injection interval $[t_{\text{start}}, t_{\text{end}}]$ published with the ROAD dataset: a benign-before region ($t < t_{\text{start}}$), the attack interval ($t_{\text{start}} \leq t \leq t_{\text{end}}$), and a benign-after region ($t > t_{\text{end}}$). Ambient captures form a single benign region. Sliding windows are then constructed within each region independently. This partitioning is used to construct and label the evaluation windows; at deployment the detector windows
the incoming stream directly.

Only windows containing exactly $\omega$ frames are retained; the trailing partial window of each region is discarded. This keeps $\omega$ consistent with its definition as the number of frames per graph. Retaining partial windows would make graph size vary systematically between regions: because each region contributes a fixed number of undersized windows, these form a larger fraction of short regions, which in turn correlates graph size with the class label.

Within each sliding window, an MSG is constructed. The MSG consists of the following components:
\begin{itemize}
    \item \textbf{Nodes:} Each node corresponds to a unique ID.
    \item \textbf{Edges:} Directed edges capture sequential relationships between IDs \cite{islam2020graph}.
\end{itemize}

MSGs capture the sequential structure of CAN message interactions by explicitly modeling transitions between IDs, enabling analysis of communication patterns across ECUs.

\subsection{Time Series Extraction and Analysis}

Concurrently with MSG creation, raw CAN messages are decoded using a DBC file. This transformation converts binary payloads into real-valued signals, yielding a time-series representation of CAN data. We use CAN-D to facilitate this extraction process \cite{verma2021can}. The resulting time-series representation captures temporal patterns and trends for each signal associated with an ID within a sliding window, which helps characterize masquerade attacks that rely on subtle deviations in signal values. These signal-level representations are later used to compute statistical features for graph annotation.

\subsection{Graph Annotation}


We enrich MSGs with signal-level context through graph annotation. Each node is augmented with statistical attributes computed from the signals carried by its ID, restricted to the time span of the window in which the node appears: for each signal we compute the mean ($\mu$) and standard deviation ($\sigma$) over the signal samples falling between the first and last frame of that window. Node feature vectors are zero-padded to a common width, determined by the maximum signal count over all IDs appearing in either the ambient or the attack capture, so that every graph in an experiment has identical feature dimensionality. A single implementation of this annotation is used for benign and malicious windows alike, ensuring node features are computed identically for both classes.

We additionally attach to each edge the frequency of the corresponding transition within the window; these frequencies are supplied to the GCN as edge weights. The annotated graph thus preserves the window's communication topology while enriching nodes and edges with signal-level and transition statistics, supporting the ability of subsequent GNN models to detect masquerade attacks.

\begin{figure*}
    \centering
    \includegraphics[width=1\linewidth]{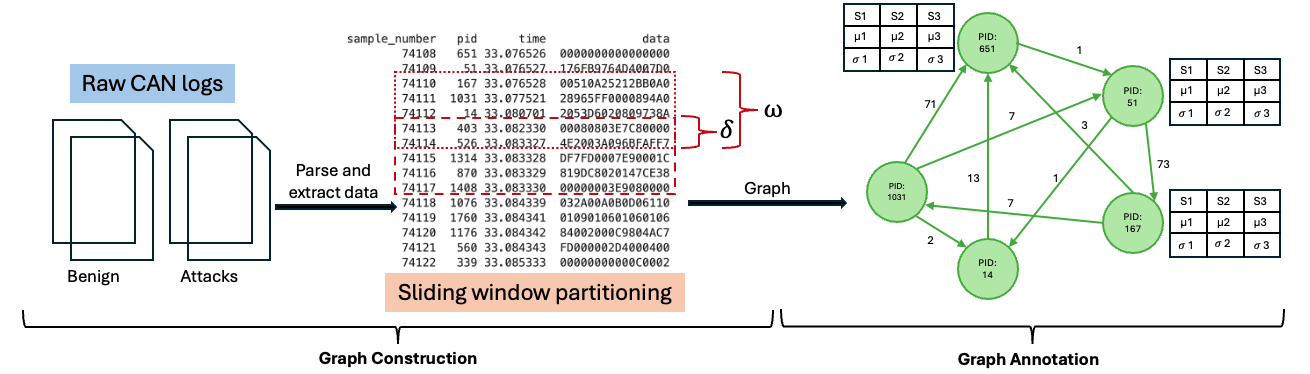}
    \caption{Parsing CAN messages to extract ID, time stamps, and signal data, then sliding window partitioning to determine the number of nodes in each graph (graph construction). Annotating with node and edge attributes (graph annotation).}
    \label{fig:steps}
\end{figure*}

Fig. \ref{fig:steps} illustrates the graph construction and graph annotation stages of the pipeline. After parsing CAN message logs into timestamps, IDs, and signal data, the data is partitioned into sliding windows. For each window, an MSG is constructed and subsequently annotated with node- and edge-level statistical features.

\subsection{Graph Classification with GNNs}
\label{sec:graph_classification}

We formulate masquerade attack detection as a graph classification problem. Each annotated MSG constructed from a sliding window is represented as a graph $\mathcal{G} = (\mathcal{V}, \mathcal{E})$ with node feature matrix $\mathbf{H}^{(0)} \in \mathbb{R}^{|\mathcal{V}| \times d}$, where $d$ denotes the dimensionality of the node features (i.e., the number of statistical attributes associated with each node, such as $\mu$ and $\sigma$).

The objective is to learn a function
\[
f: \mathcal{G} \rightarrow \{0,1\},
\]
where $0$ denotes benign behavior and $1$ denotes malicious activity. Applying this formulation to the segmented CAN stream, each sliding window yields a labeled graph $G_i$ with label $y_i \in \{0, 1\}$. Labels are derived from the ground-truth injection interval rather than from the capture a window originates from:
windows constructed within the attack interval are labeled malicious ($y_i = 1$), while windows from the benign-before and benign-after regions, and all windows from ambient captures, are labeled benign ($y_i = 0$). Because windows are constructed within a single region (Sec.V-B), no window spans the injection boundary and every frame in a window shares its label. This results in a dataset of labeled graphs $\{(\mathcal{G}_i, y_i)\}_{i=1}^N$.

The goal is to learn a function $f$, parameterized by a GNN, that generalizes to unseen graphs by minimizing a classification loss over this dataset.

To learn representations from $\mathcal{G}$, we employ a GCN. GCNs are a widely used class of GNNs that effectively capture local structural dependencies  and node features through neighborhood aggregation \cite{kipf2017semi}.  We adopt a GCN architecture due to its simplicity, computational efficiency, and strong empirical  performance as a baseline for graph-level classification tasks.

GNNs operate through a message-passing mechanism. At each layer $k$, node representations are updated by aggregating information from neighboring nodes and combining it with the  node’s current representation:
\begin{equation*}
\begin{aligned}
\mathbf{h}_u^{(k)} 
&= \mathrm{UPDATE}^{(k)}\!\Big(
    \mathbf{h}_u^{(k-1)}, \\
&\quad \mathrm{AGGREGATE}^{(k)}\!\big(
    \{\mathbf{h}_v^{(k-1)} : v \in \mathcal{N}(u)\}
\big)
\Big)
\end{aligned}
\end{equation*}
where $\mathbf{h}_u^{(k)} \in \mathbb{R}^d$ denotes the embedding of node $u$ at layer $k$, and $\mathcal{N}(u)$ is the set of neighbors of $u$. Collecting node embeddings for all nodes, we define the matrix $\mathbf{H}^{(k)} \in \mathbb{R}^{|\mathcal{V}| \times d}$, where each row corresponds to the embedding of a node at layer $k$.

In matrix form, the GCN update can be written as:
\[
\mathbf{H}^{(k)} = \sigma\!\left(
    \tilde{\mathbf{D}}^{-1/2} \tilde{\mathbf{A}} \tilde{\mathbf{D}}^{-1/2} \mathbf{H}^{(k-1)} \mathbf{W}^{(k)}
\right),
\]
where $\mathbf{A}$ is the weighted adjacency matrix of the directed graph $\mathcal{G}$, with $\mathbf{A}_{ij}$ representing the weight of the edge from node $j$ to node $i$. We define $\tilde{\mathbf{A}} = \mathbf{A} + I_{|\mathcal{V}|}$, where $I_{|\mathcal{V}|}$ is the identity matrix that introduces self-connections. The diagonal degree matrix is defined as $\tilde{\mathbf{D}}_{ii} = \sum_{j} \tilde{\mathbf{A}}_{ij}$, $\mathbf{W}^{(k)}$ is a trainable weight matrix, and $\sigma$ is a non-linear activation function \cite{goel2025community}.

To obtain a graph-level prediction, node representations from the final layer are aggregated into a global graph representation:
\[
\mathbf{z}_{\mathcal{G}} = \mathrm{AGGREGATE}\!\left(\{\mathbf{h}_u^{(K)} \mid u \in \mathcal{V}\}\right),
\]
where $\mathrm{AGGREGATE}$ denotes a permutation-invariant function over node embeddings (e.g., mean or sum), which is then used as input to a graph-level classification function:
\[
\hat{y} = f(\mathbf{z}_{\mathcal{G}}).
\]

\begin{figure*} [ht]
    \centering
    \includegraphics[width=1\linewidth]{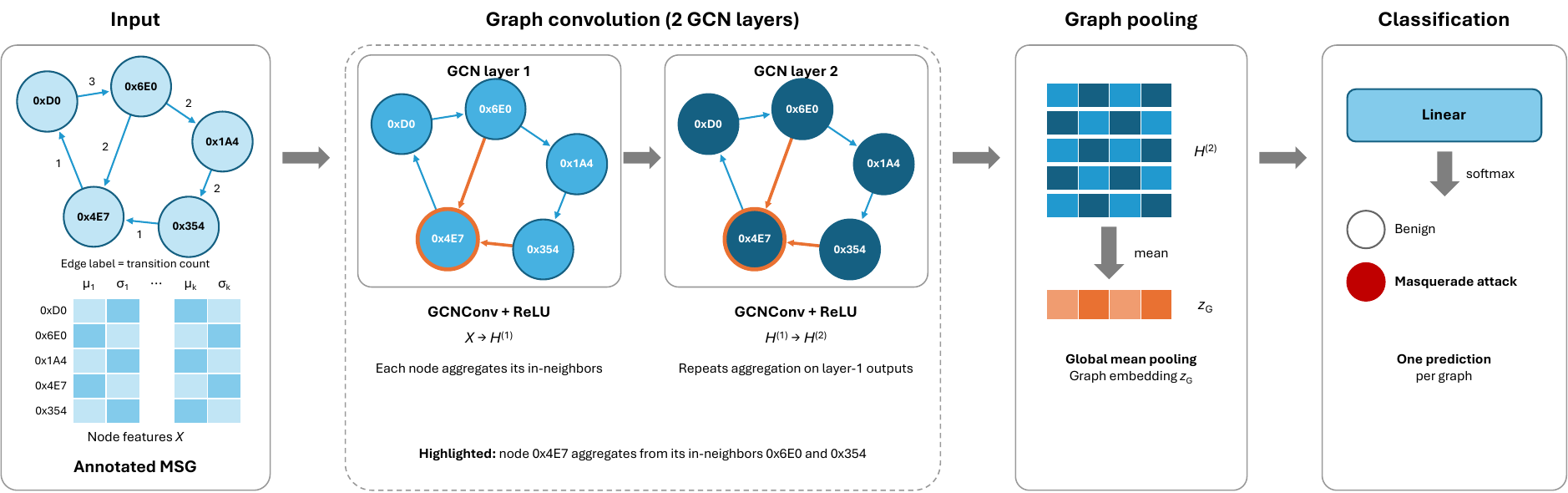}
    \caption{GCN-based graph-level classification in PhantoMasq. An annotated, directed message-sequence graph is processed by two graph convolutional layers that aggregate node information. Global mean pooling produces a graph embedding, which is used to classify each window as benign or a masquerade attack.}
    \label{fig:gcn}
\end{figure*}

The overall model thus consists of three stages:
\begin{enumerate}
    \item \textbf{Node feature aggregation:} Node features, including statistical annotations ($\mu$, $\sigma$), are propagated across GCN layers using graph connectivity and edge weights.
    \item \textbf{Global graph embedding:} An aggregation operation combines node embeddings into a graph-level representation.
    \item \textbf{Classification:} The graph embedding is used to predict whether the graph is benign or malicious.
\end{enumerate}
Fig. \ref{fig:gcn} illustrates the graph convolution process for graph-level intrusion detection.

\section{PhantoMasq Implementation}
\label{sec:implementation}

PhantoMasq was implemented in \texttt{Python 3.13.2}, using \texttt{NetworkX 3.4.2} for graph operations, \texttt{PyTorch 2.7.0} and \texttt{PyTorch Geometric (PyG) 2.6.1} to build our GNN, and \texttt{scikit-learn 1.6.1} for evaluation. Our code is available at \cite{nehal2023gnns_can}. The released code includes the full pipeline, from raw CAN data preprocessing and MSG construction to feature extraction, model training, and evaluation. All experiments were run on an Apple M4 system with a 10-core CPU and 24~GB of RAM; the GPU comparison in Section~\ref{sec:gnn_setup} used the M4's integrated GPU via PyTorch's Metal Performance Shaders (MPS) backend.

\subsection{GNN Setup}
\label{sec:gnn_setup}



We implement a GNN-based model for graph-level classification over the annotated MSGs. The model is a two-layer GCN: each layer applies GCNConv with ReLU activation, $\mathrm{ReLU}(x) = \max(x, 0)$, consuming node features, edge indices, and edge weights. Node embeddings from the final layer are aggregated by global mean pooling into a graph-level representation, where a linear layer maps to the two classes. The hidden dimensionality is 64. The model is trained with Adam (learning rate 0.01) on cross-entropy loss for 30 epochs with a batch size of 64; during training, graphs are batched using PyG.

The ambient capture and the three regions of each attack
capture are split temporally into training and test portions,
using the first 70\% and remaining 30\% of frames,
respectively. This split is performed \emph{before} windowing.
Windows are then constructed independently within each
portion, ensuring that no window spans the train/test boundary
and no frame appears in both training and test windows.
This holds for all configurations, including those with
overlapping windows ($\delta < \omega$). The resulting training and test datasets both contain benign and malicious windows.

No validation set is used. All hyperparameters, including the number of training epochs, were fixed before any results were obtained and were not adjusted in response to measured performance. Each model is trained for the full epoch budget,
and the test set is evaluated once, on the final-epoch model. Each configuration is repeated with five random seeds, and we report means across seeds.

Training and inference are performed on CPU. Inference latency is measured at batch size 1, reflecting the per-window latency of an in-vehicle deployment in which windows arrive sequentially. On our test hardware, CPU inference at batch size 1 was also faster than GPU inference, and we report CPU timings throughout.

\subsection{Dataset}

We use the Real ORNL Automotive Dynamometer (ROAD) CAN IDS dataset \cite{verma2024comprehensive}, which provides benign and attack CAN traffic collected from a real vehicle at Oak Ridge National Laboratory (ORNL). The dataset comprises over 3.5 hours of CAN traffic, including ambient data from normal driving activities and multiple simulated attack scenarios, including masquerade attacks. The data is provided in a signal-level time-series format, which supports the feature extraction and graph construction process used in this work.

The presence of labeled masquerade attack scenarios, together with signal-level data, enables systematic evaluation of graph-based intrusion detection methods under realistic signal and communication patterns. We consider five masquerade attack cases grouped into four scenario types:

\subsubsection{\textbf{Correlated Signal Attack}}

In this attack, a single ID responsible for communicating the speeds of all four wheels (encoded as a two-byte signal) is injected with four false speed values, one for each wheel.

\subsubsection{\textbf{Max Speedometer Attack}}

In this attack, the one-byte speedometer signal is replaced with the value 0xFF, causing it to report its maximum value.

\subsubsection{\textbf{Max Engine Coolant Temperature Attack}}

In this attack, the one-byte engine coolant signal is set to 0xFF, triggering a maximum temperature reading.

\subsubsection{\textbf{Reverse Lights Off Attack}}

 This attack targets a one-bit signal representing the state of the reverse lights, forcing it to remain off.

\subsubsection{\textbf{Reverse Lights On Attack}}

This attack targets the same signal, forcing it to remain on.

We use all ROAD masquerade captures: three instances each of the correlated signal, max speedometer, reverse lights off, and reverse lights on attacks, and the single max engine coolant temperature capture, for thirteen captures in total. Ambient traffic is drawn from the dyno drive (winter) capture. Per-type results average over that type's instances.

In each masquerade attack, the adversary suppresses the ECU responsible for one CAN identifier and injects spoofed frames using that same identifier at its normal transmission rate. We refer to this identifier as the \emph{targeted ID} (recorded as \texttt{injection\_id} in the ROAD metadata). Only frames bearing
the targeted ID are modified; all other identifiers on the bus transmit normally for the duration of the attack. Table~\ref{tab:targeted_ids} lists the targeted ID, its transmission rate, and the manipulated signal for each attack.

Because only the targeted ID is altered, the observable deviation within a sliding window depends on two properties of the attack: how often the targeted ID is transmitted, which determines whether its frames appear in a given window at all, and the magnitude of the change applied to its signal. Max speedometer and
both reverse-light attacks target the same identifier at the same rate, so differences among them isolate the effect of deviation magnitude. These two factors account for much of the variation in detection performance across attacks (Section~\ref{sec:detection_perf}).

\begin{table}[!htb]
    \centering
    \caption{Targeted ID and manipulated signal for each masquerade attack.
             Rates and traffic shares are measured over each capture's
             injection interval.}
    \label{tab:targeted_ids}
    \footnotesize
    \setlength{\tabcolsep}{3pt}
    \begin{tabularx}{\columnwidth}{@{}lccc>{\raggedright\arraybackslash}X@{}}
    \toprule
    \textbf{Attack} & \textbf{ID} & \textbf{Rate} & \textbf{Share} &
    \textbf{Manipulated signal} \\
    \midrule
    Correlated signal  & 0x6E0 & 100 Hz & 4.46\% & Wheel speeds (2-byte, $\times$4) \\
    Max speedometer    & 0xD0  & 100 Hz & 4.42\% & Speedometer (1-byte $\to$ 0xFF) \\
    Max engine coolant & 0x4E7 & 10 Hz  & 0.45\% & Coolant temp.\ (1-byte $\to$ 0xFF) \\
    Reverse lights off & 0xD0  & 100 Hz & 4.46\% & Reverse lights (1-bit) \\
    Reverse lights on  & 0xD0  & 100 Hz & 4.46\% & Reverse lights (1-bit) \\
    \bottomrule
    \end{tabularx}
\end{table}

\subsection{Evaluation Setup} \label{subsec:Evaluation Setup}
\label{sec:eval_setup}

Evaluation is performed at the sliding window level, with each window labeled as described in Section~\ref{sec:graph_classification}. To assess detection performance, we use the Area Under the Receiver Operating Characteristic Curve (AUC-ROC) \cite{fawcett2006introduction}, a threshold-independent metric that measures the ability of the model to distinguish between benign and malicious instances. AUC-ROC can be interpreted as the probability that a randomly chosen malicious instance is ranked higher than a randomly chosen benign instance. Mathematically, it is defined as:
\[
\mathrm{AUC\text{-}ROC}
= \int_{0}^{1} \mathrm{TPR}\!\left(\mathrm{FPR}^{-1}(x)\right)\, dx.
\]


While AUC-ROC measures how well the model distinguishes
between benign and malicious windows, it does not fully
capture detection performance when attack prevalence is low.
This is relevant to our evaluation, where the proportion of
malicious windows varies across attacks, from 5.4\% for
max engine coolant to 31.0\% for max speedometer.
A model may achieve a high AUC-ROC while exhibiting
low precision at practically relevant operating points. We therefore summarize precision--recall performance using
average precision (AP), reported as AUC-PR throughout
this paper:
\begin{equation}
    \mathrm{AUC\text{-}PR}
    = \sum_{n} (R_n - R_{n-1}) P_n ,
\end{equation}
where $P_n$ and $R_n$ denote precision and recall at the
$n$-th operating point, respectively, ordered by increasing
recall. Thus, each precision value is weighted by the
corresponding increase in recall. The baseline AUC-PR
for a random classifier equals the prevalence of the
positive class, which we report alongside it.
AUC-ROC is retained as the primary metric for consistency
with prior work~\cite{marfo2025detecting}.

In addition to detection performance, we evaluate computational efficiency using the Testing Time per Window (TTW) metric \cite{nichelini2023canova}, defined as:
\[
\mathrm{TTW} = \frac{\text{Total Detection Time}}{\text{Number of Windows}}.
\]
We also report throughput, defined as:
\[
\mathrm{Throughput} = \frac{\text{Number of Windows}}{\text{Total Detection Time}}.
\]

While TTW reflects per-window detection latency, throughput provides a complementary measure of system capacity in processing incoming data streams. 
We distinguish \emph{model-side} TTW, which measures graph
transfer and GNN inference, from \emph{window-processing} TTW,
which additionally includes MSG construction, signal-statistic
annotation, and graph conversion for an already segmented
window. Both are measured at batch size 1 on CPU. Data loading,
signal decoding, and sliding-window generation are performed
before timing and are therefore excluded. Runtime performance is reported in Section~\ref{subsec: Runtime Performance and Real-Time Feasibility}.

\section{Results}
A robust evaluation is essential for validating the proposed methodology and assessing the performance of PhantoMasq. In this section, we evaluate the model using AUC-ROC as the primary detection metric and analyze its behavior across three key dimensions: (i) detection performance across different masquerade attack scenarios, (ii) runtime efficiency and real-time feasibility, and (iii) comparison with a representative prior state-of-the-art method.

We evaluate PhantoMasq on all five ROAD masquerade attack types using every available capture per type (three each for correlated signal, max speedometer, reverse lights off and reverse lights on; one for max engine coolant). For each capture we sweep window size $\omega$ $\in$ $\{$50, \dots , 400$\}$ and offset $\delta$ $\in$ $\{$50, \dots , $\omega$$\}$, producing 36 configurations per capture and 468 experiments in total. Each configuration is run with five random seeds; we report the mean, and per-type figures average over that type's instances.

\subsection{Detection Performance and Robustness Across Attack Types}
\label{sec:detection_perf}

\begin{figure*}[!htb]
    \centering
    \includegraphics[width=1\textwidth]{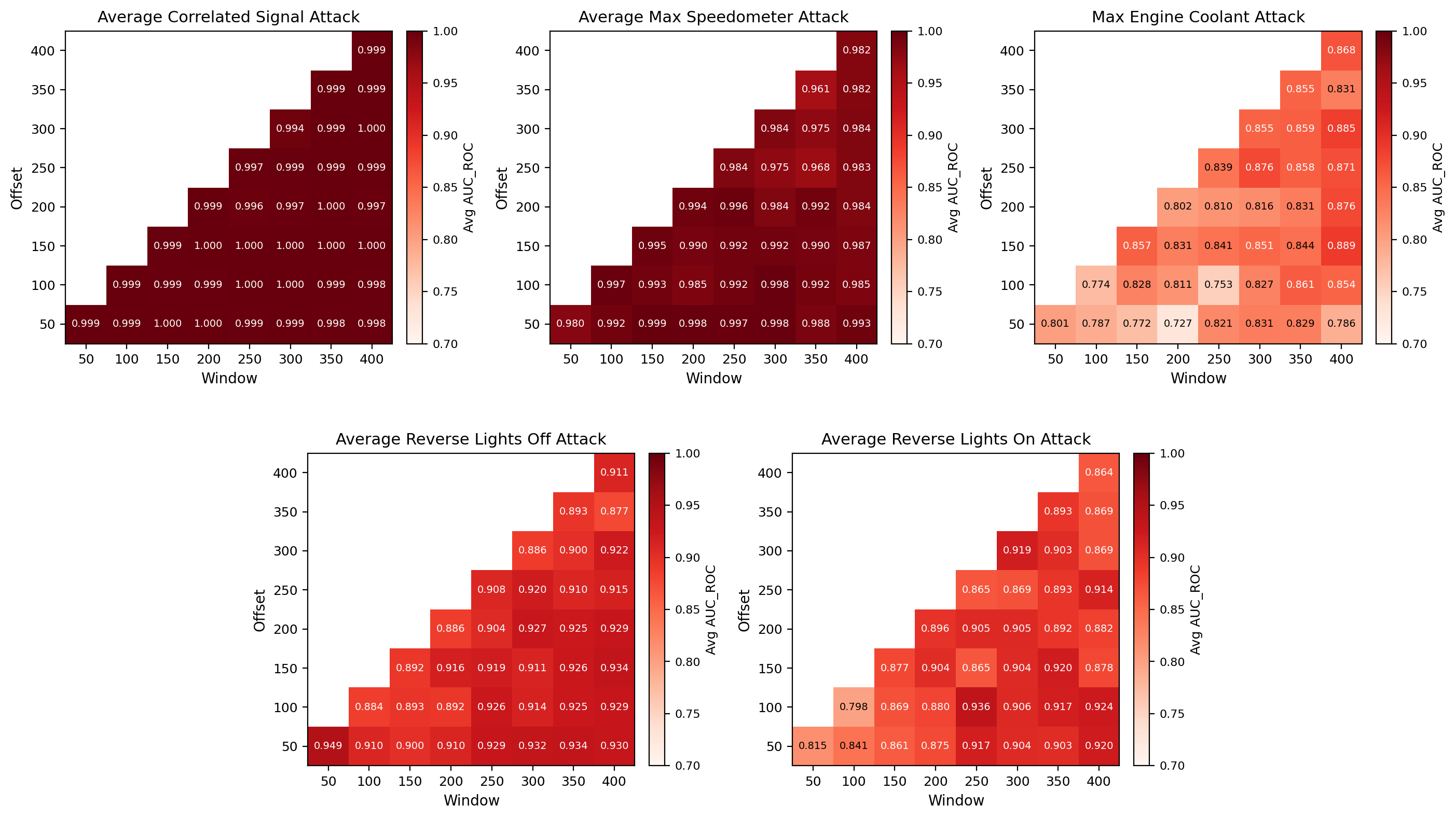}
    \caption{AUC-ROC across different $\omega$ and $\delta$ combinations for masquerade attacks detected by incorporating time series features with GCNs.}
    \label{fig:heatmaps}
\end{figure*}

Detection performance varies with two properties of the
targeted signal: 1) the magnitude of the induced deviation
and 2) the transmission rate of the targeted ID. The correlated
signal and max speedometer attacks produce large deviations
on frequently transmitted IDs (100 Hz). Specifically, the
correlated signal attack manipulates four wheel-speed values,
while the max speedometer attack forces a one-byte signal to
0xFF. Both attacks are detected almost perfectly, averaging
0.999 and 0.988 AUC-ROC across configurations, respectively.
The reverse light attacks alter only a single bit on an equally
frequent ID, resulting in lower AUC-ROC scores of 0.913
(off) and 0.888 (on). The max engine coolant attack induces
a deviation comparable to max speedometer but targets an
ID transmitted at only 10 Hz. At the smallest window size, 78\% of windows within the injection interval contain no frames from the targeted ID. Across all window configurations, the mean AUC-ROC for this attack is 0.831. The full sweep is shown in Fig.~\ref{fig:heatmaps}.

Because prevalence varies substantially across attacks (from 5.4\% of windows for max engine coolant to 31.0\% for max speedometer) we report AUC-PR alongside AUC-ROC, with prevalence stated in Table \ref{tab:trade_off}, since the chance level of AUC-PR equals the prevalence. AUC-PR is considerably more discriminating in the low-prevalence regime. For the max engine coolant attack, an AUC-ROC of 0.831 corresponds to an AUC-PR of only 0.218 against a chance level of 0.054, whereas the correlated signal attack attains 0.995 against a chance level of 0.244, close to the ceiling of the metric. The two attacks differ by 0.168 in AUC-ROC but by 0.777 in AUC-PR. Reporting AUC-ROC alone would therefore substantially overstate performance on the low-prevalence attacks, where a model may rank windows reasonably well
yet still exhibit low precision.

The difficulty ordering tracks how much evidence of the attack a window contains. The targeted ID of the correlated-signal and max-speedometer attacks transmits at 100 Hz and constitutes $\approx$ 4.5\% of bus traffic, so it appears in essentially every window. The engine-coolant attack targets a 10 Hz ID constituting 0.45\% of traffic: at $\omega$ = 50 ($\approx$ 21 ms) only 22\% of windows within the injection interval contain any frame from the targeted ID, rising to 100\% at $\omega$ = 400 ($\approx$ 178 ms). The reverse-light attacks target a frequently transmitted ID but manipulate a single bit, so the per-window statistical deviation is small even when the ID is present.

Sensitivity to the sliding-window parameters varies across
attack types. For the correlated signal and max speedometer
attacks, AUC-ROC varies by only 0.006 and 0.038,
respectively, across the 36 configurations. In both cases,
the targeted IDs are transmitted frequently and the induced
signal deviations are large, making detection relatively
insensitive to the choice of $(\omega, \delta)$.
In contrast, performance varies more substantially for
reverse lights off (0.073), reverse lights on (0.138),
and max engine coolant (0.162). For the latter, the
targeted ID appears in fewer than a quarter of the
smallest windows. These findings suggest that the choice
of window parameters becomes more important for
challenging attacks, particularly those targeting
less frequently transmitted IDs.

Relative to the baseline evaluated on the same grid, however, PhantoMasq is considerably more stable: its AUC-ROC spread across configurations is 0.006–0.162, against 0.18–0.73 for the baseline, with the largest difference on max engine coolant (0.162 vs 0.730). All model-training hyperparameters were fixed a priori, and the relatively small variation in performance across window configurations suggests that PhantoMasq does not require extensive configuration tuning to achieve competitive performance.

\begin{table*}[!htb]
    \centering
    \caption{Trade-Off between detection accuracy and runtime efficiency for selected ($\omega$, $\delta$) configurations.}
    \label{tab:trade_off}
    \setlength{\tabcolsep}{5pt}
    \begin{tabular}{lcccccccc}
    \toprule
    \textbf{Attack Type} & $\boldsymbol{\omega}$ & $\boldsymbol{\delta}$ & \textbf{AUC-ROC} & \textbf{AUC-PR} & \textbf{Prevalence} & \makecell{\textbf{Model-Side}\\\textbf{TTW (ms)}} & \makecell{\textbf{Window-processing}\\\textbf{TTW (ms)}} & \makecell{\textbf{Throughput}\\\textbf{(windows/sec)}} \\
    \midrule
    \textbf{Correlated\_Signal} & 250 & 100 & 1.000 & 1.000 & 0.244 & 0.218 & 4.64 & 216 \\
    \textbf{Max\_Speedometer} & 150 & 50 & 0.999 & 0.997 & 0.310 & 0.189 & 3.95 & 253 \\
    \textbf{Max\_Engine\_Coolant} & 400 & 150 & 0.889 & 0.299 & 0.053 & 0.255 & 5.06 & 198 \\
    \textbf{Reverse\_Lights\_Off} & 50 & 50 & 0.949 & 0.794 & 0.196 & 0.165 & 2.44 & 410 \\
    \textbf{Reverse\_Lights\_On} & 250 & 100 & 0.936 & 0.808 & 0.237 & 0.182 & 4.86 & 206 \\
    \bottomrule
    \end{tabular}
\end{table*}

\subsection{Runtime Performance and Real-Time Feasibility} \label{subsec: Runtime Performance and Real-Time Feasibility}


We evaluate runtime performance using \emph{model-side} TTW
and \emph{window-processing} TTW, as defined in
Section~\ref{subsec:Evaluation Setup}, together with throughput.
All measurements are performed on CPU at batch size 1.
No caching or batching optimizations are applied to the timed
processing stages. Model-side timing is preceded by a short
warm-up to exclude one-time initialization costs.


Table~\ref{tab:ttw_summary} summarizes window-processing
TTW across all 468 capture-configuration experiments, with
statistics computed from individual capture-level measurements
for each attack type. TTW generally increases with $\omega$
(correlation $0.910$) and is largely independent of $\delta$.
Although the highest values typically occur at $\omega=400$,
the maxima for max speedometer ($\omega=300$, $\delta=200$)
and reverse lights off ($\omega=350$, $\delta=100$) slightly
exceed their respective measurements at $\omega=400$,
consistent with measurement variability.


Despite these variations across window configurations,
window-processing TTW ranges from 2.33 to 5.94 ms
across all 468 capture-configuration experiments
(mean 4.68 ms, standard deviation 0.66 ms).
Mean latency differs by at most 0.18 ms across attack
types, while throughput ranges from 168 to 429
windows per second.


Model-side TTW ranges from 0.148 to 0.380 ms,
substantially below the corresponding window-processing
TTW. This indicates that most of the measured processing
time is spent on MSG construction, signal-statistic
annotation, and graph conversion rather than GCN inference.
\begin{table*}[!htb]
    \centering
    \caption{TTW Summary Across Attack Types With Min/Max Configurations In Milliseconds.}
    \label{tab:ttw_summary}
    \begin{tabular}{lccccc}
    \toprule
    \textbf{Attack Type} & \textbf{Mean} & \textbf{Standard Deviation} & \textbf{Median} & \textbf{Minimum} & \textbf{Maximum} \\
    \midrule
    \textbf{Correlated\_Signal} & 4.63 & 0.67 & 4.89 & 2.33 $(\omega = 50, \delta = 50)$ & 5.43 $(\omega = 400, \delta = 350)$ \\
    \textbf{Max\_Speedometer} & 4.80 & 0.67 & 5.03 & 2.48 $(\omega = 50, \delta = 50)$ & 5.94 $(\omega = 300, \delta = 200)$ \\
    \textbf{Max\_Engine\_Coolant} & 4.63 & 0.67 & 4.87 & 2.47 $(\omega = 50, \delta = 50)$ & 5.29 $(\omega = 400, \delta = 400)$\\
    \textbf{Reverse\_Lights\_Off} & 4.62 & 0.64 & 4.86 & 2.38 $(\omega = 50, \delta = 50)$ & 5.47 $(\omega = 350, \delta = 100)$ \\
    \textbf{Reverse\_Lights\_On} & 4.70 & 0.64 & 4.94 & 2.51 $(\omega = 50, \delta = 50)$ & 5.27 $(\omega = 400, \delta = 400)$\\
    \bottomrule
    \end{tabular}
\end{table*}


For descriptive purposes, Table~\ref{tab:trade_off} reports the configuration
with the highest mean test AUC-ROC for each attack, selected
retrospectively from the 36 evaluated configurations. These
configurations differ across attacks, from $\omega=50$ for
reverse lights off to $\omega=400$ for max engine coolant.
The full sweep is shown in Fig.~\ref{fig:heatmaps}. The gain over the median
configuration is modest (+0.001 to +0.057 AUC-ROC),
indicating that the selected configurations are not isolated
peaks.


\subsection{Comparison with SOTA Method} \label{subsec:Comparison with SOTA Method}

We compare PhantoMasq against a representative state-of-the-art baseline evaluated on ROAD masquerade attacks, namely the method of Marfo et al. \cite{marfo2025detecting}. This baseline is particularly relevant because it operates on the same ROAD masquerade attack scenarios and uses a similar MSG-based representation with time-series-derived node attributes, while relying on shallow graph embeddings rather than end-to-end graph learning. This comparison provides an empirical assessment of PhantoMasq relative to a graph-embedding-based detection pipeline, although the observed performance differences cannot be attributed solely to end-to-end graph representation learning.

Unlike Table~\ref{tab:ttw_summary}, which summarizes individual
capture-level measurements, the PhantoMasq TTW ranges in
Table~\ref{tab:sota} are computed after averaging across
captures of the same attack type for each window configuration.
Marfo et al.'s released artifacts include results for
sample-based windows over the identical ($\omega$, $\delta$)
grid used here, allowing a configuration-matched comparison,
subject to the differences in evaluation protocols discussed below.

On four of the five attack types, PhantoMasq's worst configuration outperforms the baseline's best. PhantoMasq is also markedly less sensitive to configuration: its AUC-ROC spread across the grid is 0.006-0.162, against 0.18–0.73 for the baseline. For max engine coolant, the baseline attains a higher maximum (0.97 vs. 0.889), although its AUC-ROC varies substantially across configurations (0.24--0.97).

The AUC-ROC comparison uses the sample-based results
released by Marfo et al., evaluated on the same
$(\omega, \delta)$ configurations as PhantoMasq.
However, the two methods use different evaluation protocols.
Marfo et al. randomly partition windows into training
and test sets, allowing temporally overlapping windows
to appear in both sets when $\delta < \omega$
\cite{marfo_pc}. In contrast, PhantoMasq partitions
each region temporally before windowing, preventing
overlap between training and test windows. These
differences should be considered when interpreting
the reported performance. Marfo et al. also apply
SMOTE only to the training data and evaluate on
the original, unmodified test set \cite{marfo_pc}.

The primary results reported by Marfo et al. use
time-based windows ($\omega = 2$--$15$~s), with
AUC-ROC ranging from 0.86 to 0.99. These windows
are substantially longer than the sample-based
windows evaluated in PhantoMasq (0.02--0.18~s).
Consequently, the results from these different
windowing regimes are not directly comparable.

Runtime cannot be compared directly. The baseline reports TTW for its time-based configuration ($\omega = 2$--$15$~s), whose windows contain roughly 10 to 700 times as many frames as PhantoMasq's, measured on different hardware (an Intel Core i9 12900HK CPU with an NVIDIA RTX 3080 Ti GPU). We therefore report PhantoMasq's latency on its own terms (Table~\ref{tab:ttw_summary}) rather than as a ratio to the baseline.



\begin{table*}[!htb]
    \centering
    \caption{Comparison of PhantoMasq and Marfo et al.
in AUC-ROC and reported TTW across masquerade attack types.}
    \label{tab:sota}
    \begin{tabular}{lcccc}
        \textbf{Attack Type} & \textbf{PhantoMasq AUC-ROC} & \textbf{Marfo et al. AUC-ROC}$^{*}$ & \textbf{PhantoMasq TTW (ms)} & \textbf{Marfo et al. TTW (ms)}$^{\dagger}$ \\ \hline
        Correlated\_Signal & 0.994 - 1.000 & 0.6000 - 0.840 & 2.37 - 5.32 & 24.00 - 6840.00 \\
        Max\_Speedometer & 0.961 - 0.999 & 0.520 - 0.700 & 2.55 - 5.37 & 5.00 - 20820.00 \\
        Max\_Engine\_Coolant & 0.727 - 0.889 & 0.240 - 0.970 & 2.47 - 5.29 & 237.00 - 4890.00 \\
        Reverse\_Lights\_Off & 0.877 - 0.949 & 0.580 - 0.810 & 2.44 - 5.19 & 27.00 - 6220.00\\
        Reverse\_Lights\_On & 0.798 - 0.936 & 0.520 - 0.720 & 2.54 - 5.22 & 5.00 - 8530.00\\ \hline
        \textbf{\textit{Overall Range}} & \textbf{\textit{0.727 - 1.000}} & \textbf{\textit{0.240 - 0.970}} & \textbf{\textit{2.37 - 5.37}} & \textbf{\textit{5.00 - 20820.00}} \\ \hline
    \end{tabular}
    \vspace{2pt}
    \begin{minipage}{\linewidth}
    \footnotesize
    $^{*}$Sample-based results from the released artifacts of~\cite{marfo2025detecting},
    on the same ($\omega$, $\delta$) grid as PhantoMasq. Their primary time-based
    configuration ($\omega = 2$--$15$~s) reports 0.86--0.99 AUC-ROC.\\
    $^{\dagger}$Values are as reported in~\cite{marfo2025detecting} for time-based
    windows on an Intel Core i9-12900HK CPU with an NVIDIA RTX 3080 Ti GPU. They are not
    directly comparable with PhantoMasq's TTW, which was measured on sample-based windows
    ($\omega = 50$--$400$ frames) on an Apple M4 CPU.
    \end{minipage}
\end{table*}


\section{Discussion}
\label{sec:discussion}

PhantoMasq detects all five ROAD masquerade attack types, with mean AUC-ROC ranging
from 0.831 for max engine coolant to 0.999 for correlated signal, and a window-processing latency of 2.33--5.94 ms
per window on CPU. Performance is not uniform
across attacks, and the variation is informative about what makes a masquerade
attack detectable at the window level.

Because a masquerade attack modifies only frames bearing the targeted ID, the
deviation observable in a window depends on two properties of the attack: how
often the targeted identifier is transmitted, and how far the manipulated signal
departs from its legitimate values. Attacks that combine a frequently transmitted
identifier with a large deviation, such as four diverging wheel-speed values or a
speedometer forced to 0xFF, are detected almost perfectly. The reverse-light
attacks target the same 100~Hz identifier as max speedometer but alter only a
single bit, and their performance is correspondingly lower. Max engine coolant
induces a large deviation, but its targeted identifier is transmitted at only
10~Hz; at the smallest window size, 78\% of windows within the injection interval
contain no frame from it at all. A window-level detector cannot flag an attack
that leaves the window unchanged, so the window must be long enough to contain
frames from the targeted identifier for the attack to be observable.

The results also show why AUC-ROC alone is an incomplete measure for this
problem. Prevalence ranges from 5.4\% to 31.0\% across attacks, and where it is
lowest the two metrics diverge sharply: for max engine coolant, an AUC-ROC of
0.831 corresponds to an AUC-PR of only 0.218. A detector facing such an attack ranks attack windows above benign ones more often than not, yet precision can remain low. Reporting AUC-PR together with
prevalence makes this visible.

Computational cost is dominated by preprocessing rather
than inference. Model-side TTW ranges from 0.15--0.38 ms per window,
while window-processing TTW, which additionally includes
MSG construction, signal-statistic annotation, and graph
conversion, ranges from 2.33--5.94 ms. Latency grows with window size and is otherwise nearly
identical across attacks. The ROAD captures carry
approximately 2,240 frames per second, so real-time
operation at $\delta=50$, the most demanding offset
in our grid, requires processing about 45 windows
per second; the lowest throughput we measured across
all configurations, 168 windows per second, exceeds
this by a factor of about 3.7. These measurements
were taken on a laptop-class CPU, and embedded
automotive processors are slower, so real-time
feasibility on in-vehicle hardware remains to be
confirmed.

Compared with the graph-embedding baseline of Marfo
et~al.~\cite{marfo2025detecting} evaluated on the same sliding-window
configurations, PhantoMasq achieves higher AUC-ROC on all five attack types and varies
far less across configurations, with an AUC-ROC spread of 0.006--0.162 against
0.18--0.73 for the baseline. One plausible explanation is end-to-end learning:
rather than fitting a separate classifier to precomputed embeddings, the GCN
learns representations directly under the detection objective. We have not
isolated this factor experimentally, and the two methods also differ in their
classifiers and in how structural and signal features are combined. Runtime is not
directly comparable, as the baseline's reported TTW was measured for time-based
windows containing roughly 10 to 700 times as many frames, on different hardware.

These results carry two practical implications for vehicular IDS design. First,
window parameters should be chosen with reference to the signals being protected:
low-rate identifiers require windows long enough to contain their frames. Second,
because detection is easiest when an attack produces large signal deviations,
evaluations on attacks that force signals to extreme values, as most ROAD
masquerade attacks do, may overstate performance against adversaries who inject
plausible values instead.

Our study has three main limitations. First, the evaluation is
restricted to the ROAD dataset~\cite{verma2024comprehensive},
with training and testing performed on different temporal
portions of the same attack captures. Benign traffic is also
drawn from a single ambient capture. Thus, the results do not
establish generalization to unseen captures, attack types,
or vehicles. Second, the evaluation excludes windows that
cross attack boundaries, which would occur during continuous
monitoring. Third, PhantoMasq relies on decoded signal-level
information and is most effective when attacks produce
observable deviations from legitimate behavior. Its
applicability may therefore be limited when signal decoding
is unavailable or when malicious values closely resemble
normal operating conditions.

Overall, PhantoMasq shows that message sequence graphs
enriched with lightweight signal statistics support accurate
and efficient detection of masquerade attacks that produce
substantial signal deviations, and that detection difficulty
varies with the transmission rate of the targeted identifier
and the magnitude of the induced deviation.

\section{Conclusion}
\label{sec:conclusion}

This paper introduces PhantoMasq, a graph-based framework for detecting masquerade
attacks in CAN systems. By modeling CAN traffic as message sequence graphs annotated
with lightweight signal statistics and formulating intrusion detection as a
graph-level classification problem, the approach learns jointly from message
transitions and signal behavior. We evaluate PhantoMasq on all ROAD masquerade captures
under a protocol that prevents overlapping windows from spanning training and test
data. Across five attack types, PhantoMasq achieves mean AUC-ROC between 0.831 and
0.999. Compared with a state-of-the-art graph-embedding baseline evaluated on the same
sliding-window configurations, it achieves higher AUC-ROC on all five attack types and
is less sensitive to window configuration.

A key finding of this work is that detection difficulty varies with the transmission
rate of the targeted identifier and the magnitude of the induced signal deviation.
Attacks that produce large deviations on frequently transmitted identifiers are
detected almost perfectly, while performance is lower for attacks that alter a single
bit or target a sparsely transmitted identifier. Prevalence also matters: for max
engine coolant, where only 5.4\% of windows are malicious, an AUC-ROC of 0.831
corresponds to an AUC-PR of only 0.218, showing that AUC-ROC alone can overstate
performance on low-prevalence attacks.

PhantoMasq is also computationally efficient. Window-processing TTW ranges from 2.33--5.94 ms per window on CPU, with throughput of 168--429 windows per second and processing cost dominated by graph construction and annotation rather than GNN inference. This throughput exceeds the rate required to keep pace with ROAD traffic on our test hardware, although real-time feasibility on embedded automotive processors remains to be confirmed.

Future work includes evaluating generalization across captures, attack types, and
datasets, and validating PhantoMasq on embedded hardware with live CAN streams.

\section*{Acknowledgments} \label{sec:Acknowledgments}

This manuscript has been authored by UT-Battelle, Limited Liability Company (LLC) under Contract No. DE-AC05-00OR22725 with the U.S. Department of Energy (DOE). The publisher, by accepting the article for publication, acknowledges that the U.S. Government retains a non-exclusive, paid-up, irrevocable, worldwide license to publish or reproduce the published form of this manuscript, or allow others to do so, for U.S. Government purposes. The DOE will provide public access to these results in accordance with the DOE Public Access Plan (http://energy.gov/downloads/doe publicaccess-plan). This research was sponsored in part by Oak Ridge National Laboratory’s Laboratory Directed Research and Development program, DOE Omni Technology Alliance Internship Program, and by the U.S. DOE. No additional external funding was received. The funders had no role in study design, data collection and analysis, decision to publish, or preparation of the manuscript.

\bibliographystyle{IEEEtran}
\bibliography{references_ieeetran_cleaned}  

@ARTICLE{Shahriar:2023:CANShield,
  author={Shahriar, Md Hasan and Xiao, Yang and Moriano, Pablo and Lou, Wenjing and Hou, Y. Thomas},
  journal={IEEE Internet Things J.}, 
  title={CANShield: Deep-Learning-Based Intrusion Detection Framework for Controller Area Networks at the Signal Level}, 
  year={2023},
  volume={10},
  number={24},
  pages={22111-22127},
  doi={10.1109/JIOT.2023.3303271}}

@inproceedings{kipf2017semi,
  title={Semi-Supervised Classification with Graph Convolutional Networks},
  author={Kipf, Thomas N. and Welling, Max},
  booktitle={International Conference on Learning Representations (ICLR)},
  year={2017}
}

@inproceedings{wang2025dynamic,
  title={Dynamic Graph-Based Semi-supervised Anomaly Detection for In-Vehicle {CAN} Bus Network},
  author={Wang, Chao and Zhang, Qiuyu and He, Yunhua and Xiao, Ke},
  booktitle={Int. Conf. Wireless Artif. Intell. Comput. Syst. Appl.},
  pages={332--342},
  year={2025}
}

@article{moriano2022detecting,
  title={Detecting {CAN} Masquerade Attacks With Signal Clustering Similarity},
  author={Moriano, Pablo and Bridges, Robert A and Iannacone, Michael D},
  journal={arXiv},
  eprint={2201.02665},
  year={2022}
}

@article{zhang2023federated,
  title={Federated graph neural network for fast anomaly detection in controller area networks},
  author={Zhang, Hengrun and Zeng, Kai and Lin, Shuai},
  journal={IEEE Trans. Inf. Forensics Security},
  volume={18},
  pages={1566--1579},
  year={2023}
}

@inproceedings{he2024d,
  title={A\&D Graph-Based Graph Neural Network Intrusion Detection for In-Vehicle Controller Area Network},
  author={He, Yaru and Gao, Jiaqi and Fan, Mingrui and Han, Daoqi and Lu, Yueming and Qiao, Yaojun},
  booktitle={Proc. IEEE/CIC Int. Conf. Commun. China (ICCC)},
  pages={1281--1286},
  year={2024}
}

@article{du2025ugl,
  title={UGL: A comprehensive hybrid model integrating {GCN} and {LSTM} for enhanced intrusion detection in {UAV} controller area networks},
  author={Du, Ying and Li, Yilong and Cheng, Pu and Han, Zhijie and Wang, Yanan},
  journal={Comput. Netw.},
  volume={262},
  pages={111157},
  year={2025}
}

@inproceedings{el2024safeguarding,
  title={Safeguarding Smart Vehicles: {GNN}-Powered Real-Time {IDS} for {CAN} Networks},
  author={El-Dawy, Sarah and Hussien, Abdallah and Ammar, Mostafa and El-Dawy, Ahmed},
  booktitle={Proc. Int. Conf. Mach. Intell. Smart Innov. (ICMISI)},
  pages={228--231},
  year={2024}
}

@article{islam2020graph,
  title={Graph-based intrusion detection system for controller area networks},
  author={Islam, Riadul and Refat, Rafi Ud Daula and Yerram, Sai Manikanta and Malik, Hafiz},
  journal={IEEE Trans. Intell. Transp. Syst.},
  volume={23},
  number={3},
  pages={1727--1736},
  year={2020}
}

@article{ye2025gdt,
  title={GDT-{IDS}: Graph-based decision tree intrusion detection system for controller area network},
  author={Ye, Pengdong and Liang, Yanhua and Bie, Yutao and Qin, Guihe and Song, Jiaru and Wang, Yingqing and Liu, Wanning},
  journal={J. Supercomput.},
  volume={81},
  number={4},
  pages={591},
  year={2025}
}

@inproceedings{yang2024anomaly,
  title={Anomaly Detection in {CAN} Bus Based on Spatio-Temporal Graph Convolutional Networks},
  author={Yang, Shuya and Yang, Dan and Qu, Junsuo and Guo, Yongxin},
  booktitle={Proc. 10th Int. Conf. Mech. Electron. Eng. (ICMEE)},
  pages={7--13},
  year={2024}
}

@article{hasan2025can,
  title={{CAN}-GraphiT: A Graph-Based {IDS} for {CAN} Networks using Transformer},
  author={Hasan, Md Mahbub and Ghose, Sanchita and Roy, Krishna Chandra},
  journal={IEEE Access},
  year={2025}
}

@article{jedh2021detection,
  title={Detection of message injection attacks onto the can bus using similarities of successive messages-sequence graphs},
  author={Jedh, Mubark and Othmane, Lotfi Ben and Ahmed, Noor and Bhargava, Bharat},
  journal={IEEE Trans. Inf. Forensics Security},
  volume={16},
  pages={4133--4146},
  year={2021}
}

@article{verma2024comprehensive,
  title={A Comprehensive Guide to {CAN} {IDS} Data and Introduction of the {ROAD} Dataset},
  author={Verma, Miki E and Bridges, Robert A and Iannacone, Michael D and Hollifield, Samuel C and Moriano, Pablo and Hespeler, Steven C and Kay, Bill and Combs, Frank L},
  journal={PLoS One},
  volume={19},
  number={1},
  pages={e0296879},
  year={2024}
}

@article{gao2025signal,
  title={Signal-relationship-aware explainable intrusion detection in controller area networks using graph transformers},
  author={Gao, Fei and Liu, Jinshuo and Li, Chengzhe and Gao, Zhenhai and Zhao, Rui},
  journal={Knowl.-Based Syst.},
  pages={114237},
  year={2025}
}

@article{goel2025community,
 title = {Community detection robustness of graph neural networks},
  author = {Goel, Jaidev and Moriano, Pablo and Kannan, Ramakrishnan and Gel, Yulia R.},
  journal = {Phys. Rev. Res.},
  volume = {8},
  issue = {2},
  pages = {023123},
  numpages = {26},
  year = {2026},
  month = {May},
  publisher = {American Physical Society},
  doi = {10.1103/v6nq-hcm8}
}

@article{zhong2024survey,
  title={A survey on graph neural networks for intrusion detection systems: Methods, trends and challenges},
  author={Zhong, Meihui and Lin, Mingwei and Zhang, Chao and Xu, Zeshui},
  journal={Comput. \& Security},
  volume={141},
  pages={103821},
  year={2024}
}

@book{hamilton2020graph,
  title={Graph Representation Learning},
  author={Hamilton, William L},
  year={2020},
  publisher={Morgan \& Claypool Publishers}
}

@inproceedings {191966,
author = {Ian Foster and Andrew Prudhomme and Karl Koscher and Stefan Savage},
title = {Fast and Vulnerable: A Story of Telematic Failures},
booktitle = {9th USENIX Workshop on Offensive Technologies (WOOT 15)},
year = {2015},
address = {Washington, D.C.},
url = {https://www.usenix.org/conference/woot15/workshop-program/presentation/foster},
publisher = {USENIX Association},
month = aug
}

@article{miller2015remote,
  title={Remote Exploitation of an Unaltered Passenger Vehicle},
  author={Miller, Charlie and Valasek, Chris},
  journal={Black Hat USA},
  volume={2015},
  number={S 91},
  pages={1--91},
  year={2015}
}

@article{bbc2015jeep,
  author       = {{BBC News}},
  title        = {Fiat Chrysler Recalls 1.4m Cars After Jeep Hack},
  journal      = {BBC News},
  year         = {2015},
  month        = {July},
  day          = {24},
  url          = {https://www.bbc.com/news/technology-33650491},
  note         = {Accessed: Dec. 16, 2025}
}

@article{guardian2016tesla,
  author       = {{The Guardian}},
  title        = {Tesla Model S Hacked Remotely by Chinese Security Researchers},
  journal      = {The Guardian},
  year         = {2016},
  month        = {September},
  day          = {20},
  url          = {https://www.theguardian.com/technology/2016/sep/20/tesla-model-s-chinese-hack-remote-control-brakes},
  note         = {Accessed: Dec. 16, 2025}
}

@inproceedings{burakova2016truck,
  title={Truck hacking: An experimental analysis of the $\{$SAE$\}$ j1939 standard},
  author={Burakova, Yelizaveta and Hass, Bill and Millar, Leif and Weimerskirch, Andr{\'e}},
  booktitle={Proc. 10th USENIX Workshop Offensive Technol. (WOOT)},
  year={2016}
}

@article{securityweek2023can,
  author       = {{SecurityWeek News Staff}},
  title        = {Thieves Use CAN Injection Hack to Steal Cars},
  journal      = {SecurityWeek},
  year         = {2023},
  month        = {April},
  day          = {3},
  url          = {https://www.securityweek.com/thieves-use-can-injection-hack-to-steal-cars/},
  note         = {Accessed: Dec. 16, 2025}
}

@incollection{johansson2005vehicle,
  title={Vehicle applications of controller area network},
  author={Johansson, Karl Henrik and T{\"o}rngren, Martin and Nielsen, Lars},
  booktitle={Handbook of networked and embedded control systems},
  pages={741--765},
  year={2005}
}

@article{farsi1999overview,
  title={An overview of controller area network},
  author={Farsi, Mohammad and Ratcliff, Karl and Barbosa, Manuel},
  journal={Comput. Control Eng. J.},
  volume={10},
  number={3},
  pages={113--120},
  year={1999}
}

@article{hpl2002introduction,
  title={Introduction to the controller area network ({CAN})},
  author={Steve Corrigan},
  journal={Application Report SLOA101},
  pages={1--17},
  year={2002}
}

@article{boland2021overview,
  title={An overview of can-bus development, utilization, and future potential in serial network messaging for off-road mobile equipment},
  author={Boland, Hannah M and Burgett, Morgan I and Etienne, Aaron J and Stwalley III, Robert M},
  journal={Technol. Agric.},
  year={2021}
}

@article{jo2019mauth,
  title={MAuth-{CAN}: Masquerade-attack-proof authentication for in-vehicle networks},
  author={Jo, Hyo Jin and Kim, Jin Hyun and Choi, Hyon-Young and Choi, Wonsuk and Lee, Dong Hoon and Lee, Insup},
  journal={IEEE Trans. Veh. Technol.},
  volume={69},
  number={2},
  pages={2204--2218},
  year={2019}
}

@article{jiang2025xiphos,
  title={XIPHOS: Adaptive In-Vehicle Intrusion Detection via Unsupervised Graph Contrastive Learning},
  author={Jiang, Qiguang and Wang, Kai and Wei, Yuliang and Liu, Hongri and Wang, Bailing},
  journal={IEEE Trans. Inf. Forensics Security},
  volume={20},
  pages={10419--10433},
  year={2025}
}

@inproceedings{refat2022lightweight,
  title={A lightweight intrusion detection system for can protocol using neighborhood similarity},
  author={Refat, Rafi Ud Daula and Elkhail, Abdulrahman Abu and Malik, Hafiz},
  booktitle={Proc. 7th Int. Conf. Data Sci. Mach. Learn. Appl. (CDMA)},
  pages={121--126},
  year={2022}
}

@inproceedings{meng2023gb,
  title={Gb-ids: An intrusion detection system for can bus based on graph analysis},
  author={Meng, Yan and Li, Jiachun and Liu, Fazhong and Li, Shaofeng and Hu, Haotian and Zhu, Haojin},
  booktitle={2023 IEEE/CIC International Conference on Communications in China (ICCC)},
  pages={1--6},
  year={2023}
}

@article{fawcett2006introduction,
  title={{An Introduction to {ROC} Analysis}},
  author={Fawcett, Tom},
  journal={Pattern Recognit. Lett.},
  volume={27},
  number={8},
  pages={861--874},
  year={2006}
}

@inproceedings{halder2020coids,
  title={COIDS: A clock offset based intrusion detection system for controller area networks},
  author={Halder, Subir and Conti, Mauro and Das, Sajal K},
  booktitle={Proc. 21st Int. Conf. Distrib. Comput. Netw.},
  pages={1--10},
  year={2020}
}

@inproceedings{shahriar2023cantropy,
  title={CANtropy: Time series feature extraction-based intrusion detection systems for controller area networks},
  author={Shahriar, Md Hasan and Lou, Wenjing and Hou, Y Thomas},
  booktitle={Proc. of Symp. on Vehicles Security and Privacy (VehicleSec)},
  pages={1--8},
  year={2023}
}

@inproceedings{ameen2025wavesleuth,
  title={WaveSleuth: Retrospective {PLC} Memory for Anomaly Detection in Industrial Control Systems},
  author={Ameen, Nehal and Vijayakanthan, Ramyapandian and Ayub, Adeen and Ali-Gombe, Aisha and Ahmed, Irfan},
  booktitle={Proc. IEEE Int. Symp. Hardware Oriented Security and Trust ({HOST})},
  pages={170--181},
  year={2025}
}

@inproceedings{kalle2019clik,
  title={CLIK on PLCs! Attacking control logic with decompilation and virtual {PLC}},
  author={Kalle, Sushma and Ameen, Nehal and Yoo, Hyunguk and Ahmed, Irfan},
  booktitle={Proc. Binary Anal. Res. (BAR) Workshop, {NDSS}},
  pages={1--12},
  year={2019}
}

@inproceedings{sharmin2023comparative,
  title={Comparative evaluation of anomaly-based controller area network ids},
  author={Sharmin, Shaila and Mansor, Hafizah and Abdul Kadir, Andi Fitriah and A. Aziz, Normaziah},
  booktitle={Proc. 12th Int. Conf. Softw. Comput. Appl.},
  pages={218--226},
  year={2023}
}

@book{di2012understanding,
  title={Understanding and using the controller area network communication protocol: theory and practice},
  author={Di Natale, Marco and Zeng, Haibo and Giusto, Paolo and Ghosal, Arkadeb},
  year={2012},
  publisher={Springer Science \& Business Media}
}

@article{gazdag2023crysys,
  title={CrySyS dataset of {CAN} traffic logs containing fabrication and masquerade attacks},
  author={Gazdag, Andr{\'a}s and Ferenc, Rudolf and Butty{\'a}n, Levente},
  journal={Sci. Data},
  volume={10},
  number={1},
  pages={903},
  year={2023}
}

@misc{nehal2023gnns_can,
  author       = {Ameen, Nehal and Moriano, Pablo and Ahmed, Irfan},
  title        = {{PhantoMasq}: Code and Results for "Phantom of the Masquerade"},
  year         = {2026},
  howpublished = {\url{https://github.com/nehal-vcu/PhantoMasq}}
}

@article{shen2025dbcsec,
  title={DBCSec: {{DBC}} File-Guided Secure Communication Mechanism for {{CAN}-FD} Bus},
  author={Shen, Yun and Cui, Jie and Zhong, Hong and Zhang, Jing and Zhang, Qingyang and Wei, Lu and He, Debiao},
  journal={IEEE Trans. Dependable Secure Comput.},
  year={2025}
}

@inproceedings{wu2019simplifying,
  title={Simplifying graph convolutional networks},
  author={Wu, Felix and Souza, Amauri and Zhang, Tianyi and Fifty, Christopher and Yu, Tao and Weinberger, Kilian},
  booktitle={Proc. Int. Conf. Mach. Learn. (ICML)},
  pages={6861--6871},
  year={2019}
}

@article{verma2021can,
  title={{CAN}-D: A modular four-step pipeline for comprehensively decoding controller area network data},
  author={Verma, Miki E and Bridges, Robert A and Sosnowski, Jordan J and Hollifield, Samuel C and Iannacone, Michael D},
  journal={IEEE Trans. Veh. Technol.},
  volume={70},
  number={10},
  pages={9685--9700},
  year={2021}
}

@inproceedings{cahuantzi2023comparison,
  title={A comparison of {LSTM} and {GRU} networks for learning symbolic sequences},
  author={Cahuantzi, Roberto and Chen, Xinye and G{\"u}ttel, Stefan},
  booktitle={Proc. Sci. Inf. Conf.},
  pages={771--785},
  year={2023}
}

@inproceedings{pudikov2020comparison,
  title={Comparison of {LSTM} and {GRU} recurrent neural network architectures},
  author={Pudikov, Anton and Brovko, Alexander},
  booktitle={Proc. Int. Sci. Pract. Conf. Control Eng. Decis. Making},
  pages={114--124},
  year={2020}
}

@article{marfo2025detecting,
  title={Detecting masquerade attacks in controller area networks using graph machine learning},
  author={Marfo, William and Moriano, Pablo and Tosh, Deepak K and Moore, Shirley V},
  journal={IEEE Trans. Inf. Forensics Security},
  volume={20},
  number={},
  pages={13127-13142},
  publisher={IEEE},
  year={2025}
}

@article{moriano2026evaluating,
  title={Evaluating lightweight unsupervised online {IDS} for masquerade attacks in {CAN}},
  author={Moriano, Pablo and Hespeler, Steven C and Li, Mingyan and Bridges, Robert A},
  journal={J. Inf. Secur. Appl.},
  volume={98},
  pages={104392},
  year={2026}
}

@article{nichelini2023canova,
  title={CANova: A hybrid intrusion detection framework based on automatic signal classification for {CAN}},
  author={Nichelini, Alessandro and Pozzoli, Carlo Alberto and Longari, Stefano and Carminati, Michele and Zanero, Stefano},
  journal={Comput. \& Security},
  volume={128},
  pages={103166},
  year={2023}
}

@misc{al2024can,
  title={{CAN}-MIRGU: a comprehensive {CAN} bus attack dataset from moving vehicles for intrusion detection system evaluation.},
  author={AL-KADRI, MO},
  year={2024}
}

@misc{marfo_pc,
  author       = {W. Marfo},
  howpublished = {private communication},
  month        = sep,
  year         = {2026}
}

\vfill

\end{document}